\documentclass{aa}
\usepackage{graphicx}
\usepackage{xcolor}
\usepackage{txfonts}
\usepackage{makecell}
\usepackage{hyperref}
\hypersetup{
    colorlinks=true,
    linkcolor=blue,
    filecolor=magenta,      
    urlcolor=blue,
    citecolor=blue,
    pdfpagemode=FullScreen,
    }
\usepackage{graphicx}	% Including figure files
\usepackage{amsmath}	% Advanced maths commands
\usepackage{caption}
\usepackage{subcaption}
\usepackage{multirow}
\usepackage{pifont}% http://ctan.org/pkg/pifont
\usepackage[normalem]{ulem}
\usepackage{booktabs}
\newcommand{\mass}{\log(M_{\star}/M_{\odot})}

\defcitealias{Bruzual2003}{BC16}

\begin{document}

   \title{Cosmic chronometers with galaxy clusters: a new avenue for multi-probe cosmology}

   \author{E. Tomasetti\inst{\ref{unibo},\ref{inafbo}} \fnmsep\thanks{\email{elena.tomasetti2@unibo.it}}
          \and
           M. Moresco\inst{\ref{unibo},\ref{inafbo}}
           \and
           G. Granata\inst{\ref{uferrara},\ref{unimi},\ref{icg}}%\orcidlink{0000-0002-9512-3788}
           \and
           M. D'Addona\inst{\ref{inafcapodimonte}}
           \and
           P. Bergamini\inst{\ref{inafbo}}
           \and
           C.~Grillo\inst{\ref{unimi},\ref{inafmilano}}%\orcidlink{0000-0002-5926-7143}
           \and
           A. Mercurio\inst{\ref{inafcapodimonte},\ref{usalerno},\ref{infnsalerno}}
           \and
           P.~Rosati\inst{\ref{uferrara},\ref{inafbo}}
           \and
           A. Cimatti\inst{\ref{unibo},\ref{inafbo}}
           \and
           L. Tortorelli\inst{\ref{unimunchen}}
           \and
           S. Schuldt\inst{\ref{unimi},\ref{inafmilano}}
           \and
           M. Meneghetti\inst{\ref{inafbo}}
           }

   \institute{
            Dipartimento di Fisica e Astronomia ``Augusto Righi'' -- Universit\`a di Bologna, via Piero Gobetti 93/2, I-40129 Bologna, Italy\label{unibo}
            \and
            INAF -- Osservatorio di Astrofisica e Scienza dello Spazio di Bologna, via Piero Gobetti 93/3, I-40129 Bologna, Italy\label{inafbo}
            \and
            Dipartimento di Fisica e Scienze della Terra, Universit\`a degli Studi di Ferrara, via Saragat 1, 44122 Ferrara, Italy\label{uferrara}
            \and
            Dipartimento di Fisica, Universit\`a degli Studi di Milano, via Celoria 16, I-20133 Milano, Italy \label{unimi}
            \and
            Institute of Cosmology and Gravitation, University of Portsmouth, Burnaby Rd, Portsmouth PO1 3FX, UK \label{icg}
            \and
            INAF -- Osservatorio Astronomico di Capodimonte, Via Moiariello 16, I-80131 Napoli, Italy\label{inafcapodimonte}
            \and    
            INAF -- IASF Milano, via A.~Corti 12, I-20133 Milano, Italy \label{inafmilano}
            \and
            Università di Salerno, Dipartimento di Fisica ``E.R. Caianiello'', Via Giovanni Paolo II 132, 84084 Fisciano (SA), Italy \label{usalerno}
            \and
            INFN - Gruppo Collegato di Salerno - Sezione di Napoli \label{infnsalerno}
            \and
            Universitäts-Sternwarte, Fakultät für Physik, Ludwig-Maximilians-Universität München, Scheinerstr. 1, 81679 München, Germany \label{unimunchen}
             }

    \titlerunning{Cosmic chronometers with galaxy clusters}
    \authorrunning{E. Tomasetti et al.}

\abstract{We provide a new measurement of the expansion history of the Universe at $z=0.54$ with the cosmic chronometers (CC) method, exploiting the high-quality spectroscopic VLT/MUSE data for three galaxy clusters in close-by redshift bins: SDSS J2222+2745 ($z=0.49$), MACS J1149.5+2223 ($z=0.54$), and SDSS J1029+2623 ($z=0.59$). The central one, MACS J1149.5+2223, hosts the well-known supernova ‘Refsdal’, which allowed for $H_0$ measurements via time delay cosmography (TDC). This represents the first step for a self-consistent probe combination, where different methods are applied to the same data sample. 
After selecting the most passive and massive cluster members (38 CCs), we derive their age and physical parameters via full spectrum fitting. We use the code \texttt{Bagpipes}, specifically modified to remove the cosmological prior on ages. On average, the CC sample shows super-solar metallicities $Z/Z_{\odot} = 1.3 \pm 0.7$, low dust extinction $A_{\rm{V}} = 0.3 \pm 0.3$ mag and to have formed in short bursts $\tau = 0.6 \pm 0.2$ Gyr. We also observe both an ageing trend in redshift and a mass--downsizing pattern. From the age--redshift trend, implementing the CC method through a bootstrap approach, we derive a new $H(z)$ measurement: $H$($z$=0.542) = $66_{-29}^{+81}$ (stat) $\pm$13 (syst) km/s/Mpc. We also simulate the impact of increased statistics and extended redshift coverage, finding that $H$($z$) uncertainties can be reduced by up to a factor of 4 with $\sim$100 CCs and a slightly broader redshift range (d$z\sim$0.2).}


   \maketitle
%
%-------------------------------------------------------------------

\section{Introduction}
\label{sec:1Intro}
Modern cosmology is established within the $\Lambda$CDM framework, a paradigm that has been consolidated by a set of primary cosmological probes, such as the cosmic microwave background \citep[CMB,][]{Bennett2003,PlanckCollaboration2020}, the distance ladder approach with type Ia supernovae \citep{Riess1998,Perlmutter1999}, and baryon acoustic oscillations \citep{Eisenstein2005}. Nevertheless, recent high-precision measurements have revealed a significant tension, at the 4-5$\sigma$ level, between early- and late-Universe determinations of cosmological parameters, in particular the Hubble constant \citep[e.g.][]{Kamionkowski2023,di_valentino_cosmoverse_2025}. Understanding whether this discrepancy is due to observational systematics or instead signals new physics is one of the central challenges in modern cosmology.

To address this issue, it is essential to expand the set of cosmological probes, moving beyond the traditional ones. In recent years, several emerging probes have been developed, providing new and independent ways to test the concordance model \citep[see][for a review]{Moresco2022}. Among them, the cosmic chronometers (CC) approach \citep{Jimenez2002} offers a direct way to measure the Hubble parameter, $H(z)$, through the differential ageing of passively evolving galaxies. This technique relies on minimal assumptions, and has been successfully applied up to redshift $z\simeq$ 2. At the same time, time-delay cosmography (TDC) with strong gravitational lensing has emerged as an independent and complementary probe, estimating cosmological distances through the relative time delays between the multiple images of background, time-variable, lensed sources \citep[e.g.][]{grillo_cosmography_2024,suyu_cosmology_2025,tdcosmo_collaboration_tdcosmo_2025}.

A particularly promising strategy is to combine CC and TDC within overlapping observational fields, as first suggested in \citet{Bergamini2024}. While CCs provide measurements of the expansion rate $H(z)$, TDC is sensitive to integrated distances, and their complementarity can significantly strengthen cosmological constraints. In this context, galaxy clusters play a dual role: on the one hand, massive clusters act as lenses suitable for TDC; on the other, the same clusters, together with other clusters at similar redshift, host populations of massive, passive galaxies that can serve as CCs.

In this work, we pave the way for a future synergy by applying the CC method probing the member galaxies of MACS~J1149.5+2223 \citep[][]{lotz_frontier_2017}, the strong-lensing cluster hosting the multiply imaged SN ‘Refsdal’ \citep{kelly_multiple_2015}, and clusters located in close-by redshift bins. Thanks to the SN Refsdal, MACS J1149.5+2223 enabled the first measurement of $H_0$ via TDC applied to a lensed supernova \citep{grillo_measuring_2018,grillo_accuracy_2020,kelly_magnificent_2023}, with a statistical+systematic relative uncertainty that reached approximately 6\% \citep{grillo_cosmography_2024} in a general cosmological model. By deriving independent expansion-rate constraints applying the CC approach to the cluster members, we prepare the ground for a future joint CC+TDC analysis. Such a combination holds the potential to mitigate systematic uncertainties and provide a more robust determination of the Hubble constant and the expansion history of the Universe.

Looking ahead, upcoming wide-field surveys will be transformative in this context, like LSST \citep{ivezic_lsst_2019}, which is expected to increase the supernova detection by a factor of $\sim100$ \citep{petrecca_recovered_2024}, or \textit{Euclid} \citep{Laureijs2011}. In particular, the latter is expected to revolutionize the census of strong gravitational lenses, delivering orders of magnitude more systems than currently known \citep{euclid_collaboration_euclid_2025}. This unprecedented sample will greatly expand the applicability of TDC, further enhancing the prospects for a joint CC+TDC approach within the same structures. 

This paper is organised as follows. In Sect. \ref{sec:2DATA}, we present the data analysed and the selection process followed to identify CCs. In Sect. \ref{sec:3METHOD}, we describe the full spectrum fitting technique adopted to measure the physical properties of the sample. In Sect. \ref{sec:4COSMO}, the cosmological outcome of this study is presented, and in Sect. \ref{sec:5CONCLUSIONS} we draw our conclusions.

%##############se#######  DATA #####################

\begin{figure*}[h!]
    \centering
    \includegraphics[width=0.95\linewidth]{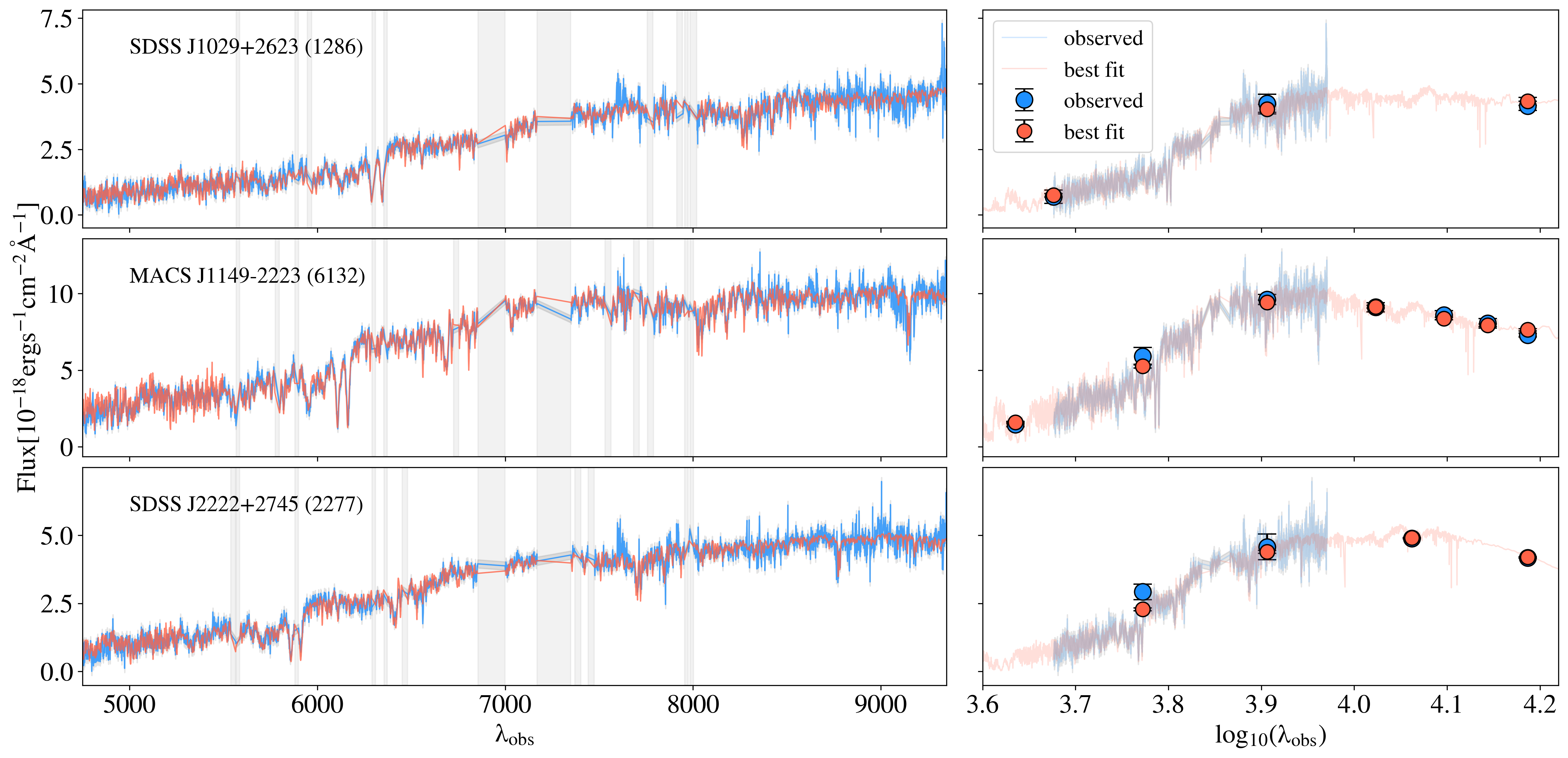}
    \caption{Fit of the spectrum (left) and photometry (right) of an example galaxy for each cluster. In blue, the observed spectrum and photometry are shown, in red the corresponding best fit. The shaded regions mark potential emission or telluric lines that are masked in the fit.}
    \label{fig:fit_example}
\end{figure*}

\section{Data}
\label{sec:2DATA}
In this section, we describe the sample under analysis, the specifics of the spectral and photometric observations, and the selection process adopted to identify our CC sample.

\subsection{Spectra and photometry}
We analysed spectra and photometry of the cluster members observed with the VLT/MUSE instrument in MACS J1149.5+2223 \citep[$z=0.54$,][]{Grillo2016,Schuldt2024}, SDSS J1029+2623 \citep[$z=0.59$,][]{acebron_vltmuse_2022}, and SDSS J2222+2745 \citep[$z=0.49$,][]{acebron_new_2022}. From here on, we will refer to them as MACS 1149, SDSS 1029 and SDSS 2222, respectively.

For all three clusters, we complemented the spectroscopic information with archival HST multicolour imaging from the ACS and the WFC3. Specifically, photometry in F475W, F814W, and F160W filters was available for SDSS 1029; in F606W, F814W, F110W, and F160W  for SDSS 2222; in F435W, F606W, F814W, F105W, F125W, F140W, and F160W for MACS 1149. Further details on the photometry are provided in Appendix \ref{appendix_observations}. For each cluster, we extracted a catalogue of structural parameters using \texttt{Morphofit}~\citep{Tortorelli2023a}, including position, velocity dispersion, and signal-to-noise ratio (\textit{S/N}). 

We weighted the MUSE cube with the members' surface brightness in the HST F814W band image, degraded and re-binned to the PSF and pixel-scale of the MUSE observations, and extracted the spectra from large circular apertures with a $1.5''$ radius, centred on the galaxy centre of light. The spectra resulting from the weighted average are representative of the central regions of the members, which are sampled with a higher weight due to their higher surface brightness. For instance, we found the velocity dispersion value obtained from the weighted spectra to be, on average, equivalent to those measured within the half-light radius of the galaxies \citep{Granata2025}. We could thus probe the central regions of the cluster galaxies with high \textit{S/N} and without the need for aperture corrections. In addition, all galaxies were inspected to ensure that the selected ones are not affected by contamination from nearby objects.\\ 
The observed spectra cover a large wavelength range, 4750 -- 9350 {\AA}, and offer high spectral resolution (R $\sim$ 3000) and \textit{S/N}. A detailed description of the observations is reported in Appendix \ref{appendix_observations}. Stellar velocity dispersions, $\sigma_{\star}$, were measured by fitting the observed spectra with pPXF \citep{cappellari_full_2023}, using selected stellar spectra from the X-shooter Spectral Library DR2 \citep{Gonneau2020} as templates. In addition, since for a few objects, the error spectrum derived from the variance data-cube appeared to be overestimated, we treated the residuals between the observed spectrum and the best-fit from pPXF as the actual noise in our spectrum. Accordingly, for each of these spectra, we adopted the standard deviation of the residuals as its $1\sigma$ uncertainty.

Lastly, in preparation for performing full spectrum fitting (FSF) jointly on spectra and photometry for each galaxy, we applied a preliminary rescaling to match the spectra with the observed photometry. Specifically, we integrated each spectrum over the F814W filter transmission curve -- available for all galaxies -- and derived a scaling factor as the ratio between the observed F814W photometric flux and this integrated value. This factor was then applied to rescale each spectrum.

\subsection{Selection of cosmic chronometers}
A critical step in applying the CC approach is the selection of a robust sample of very massive, passive galaxies with no signs of residual star formation. This ensures one can identify a homogeneous population whose star formation ceased well before the epoch of observation. Different methods have been adopted for selecting passive galaxies \citep[for a detailed review, see][]{Moresco2022}, based on their colours \citep[e.g. UVJ][]{Williams2009}, on the star formation rate \citep[SFR, e.g.][]{Pozzetti2010} or on emission lines \citep[e.g.][]{Mignoli2009}.

Concerning the sample considered here, \citet{Bergamini2024} had already identified a sample of CC candidates for each cluster, based on their stellar velocity dispersion values ($\sigma_\star > 180$ km/s) -- a stellar mass proxy in dispersion-dominated galaxies, like ellipticals -- and on their high \textit{S/N} (\textit{S/N}$>$15 in the range 3700--5250 {\AA}). Analysing the stacked spectra for each cluster, they observed that these galaxies were indeed showing the characteristics of very passive objects, owing to the red continuum, the absence of emission lines, and the presence of prominent absorption features. Here, we followed this approach by adopting the same cut in \textit{S/N}, but we adopted a less conservative cut in velocity dispersion, selecting objects with $\sigma_\star>150$ km/s to increase the statistics. With these criteria, we identified 7, 25, and 8 galaxies for SDSS 2222, MACS 1149, and SDSS 1019, respectively, totalling 40 CC candidates. The different numerosity among clusters primarily reflects the higher mass and richer galaxy population of MACS 1149.

Selecting CCs inside galaxy clusters, though, requires particular care, because of the potential environmental effects that a denser environment could have on the galaxies' evolution. The higher probability of minor and major mergers, or close encounters, can perturb systems that would otherwise evolve passively, inducing rejuvenation effects. In this sense, an effective diagnostic in removing galaxies showing even small fractions of younger components is the ratio of the CaII H, at 3969 {\AA}, and the CaII K, at 3934 {\AA} (H/K, hereafter). The first one overlaps with the H$\varepsilon$ line of the Balmer series, which is known to shrink as the stellar population ages. On the contrary, the two calcium lines tend to deepen as the population gets older, making the H/K quantity an excellent diagnostic to identify a possible residual young component in the galaxy. The typical threshold used to select CCs is H/K $<$1.2--1.5 \citep{Borghi2022a, Moresco2018}.\\
We evaluated H/K by measuring the involved pseudo-Lick indices via the public code \texttt{PyLick} \citep{Borghi2022a}. Here, we opted for a stringent limit on this feature, selecting only galaxies with H/K $<$ 1.2, which excluded just two more objects, thereby demonstrating the very low contamination present in the initial sample.

In the end, our final CC sample, covering the redshift range 0.49 -- 0.59, counted 38 massive and passive galaxies.

%%%%%%%%%%%%%%%  METHOD #########################

\section{Method and analysis}
\label{sec:3METHOD}

To benefit from the high quality of the spectro-photometric data available for this sample, we measured the physical properties of each object via FSF, jointly fitting spectra and photometry.

\subsection{Full spectrum fitting}\label{FSF}

We employed the \texttt{Bagpipes} code \citep{Carnall2018,Carnall2019} which allowed us to perform FSF with a Bayesian approach. In particular, \texttt{Bagpipes} can fit observed spectra and/or photometry with synthetic models, generated based on a set of parameters, and identify the best-fit values for these parameters by maximising the posterior probability sampled via \texttt{MultiNest} \citep{Buchner2016}, a nested sampling algorithm. 

The main ingredients required to generate the synthetic spectra are: the stellar population synthesis (SPS) models -- here the 2016 version of the \citet{Bruzual2003} models, where $\alpha$-enhancement is fixed to solar value\footnote{It will be important in future works to assess the impact of adopting models with different levels of $\alpha$-enhancement, which could lead to younger inferred ages. However, differential ages should remain largely unaffected, provided that the sample is homogeneous.}, and a \citet{Kroupa2001} initial mass function is assumed; a functional form for the star formation history (SFH); a transmission curve for the neutral interstellar medium, to account for the absorption and emission of dust \citep[e.g.][]{Calzetti2000,Cardelli1989}. In addition to these physical components, the code also allows for the inclusion of a noise parameter to account for potential underestimations of the uncertainty and a calibration component to solve possible mismatches between spectra and photometry. 

In this work, we adopted a delayed exponentially declining (DED) SFH, that has a SFR described by the functional form:
\begin{equation}\label{eq:3delayedSFH}
        {\mathrm{SFR}({t})} \propto \begin{cases} 
        {({t-T_0})}\: {\mathrm{e}^{-\frac{{t-T_0}}{\tau}}}, & t > T_0 \\ 0, & t < T_0 
        \end{cases},
\end{equation}
where $\tau$ correlates with the SFH length, and $T_0$ corresponds to the age of the Universe at which the star formation begins. This allows for a more realistic reproduction of the SFH with respect to a single burst, while keeping under control the possible degeneracies arising between the SFH characteristics and the physical parameters involved in the fit.\\
For the dust component, we adopted the \citet{Calzetti2000} law, depending only on the reddening in the V-band, $A_{\rm{V}}$. We also tested the more complex \citet{Salim2018} curve, but given the similarity in the results obtained, we opted for the simpler model.\\
We also included the noise and calibration components, the first acting as a constant multiplicative factor of the error spectrum, the other represented by a second-order Chebyshev polynomial multiplied to the whole spectrum to better match the photometry. The calibration component is also fundamental to correct for potential flux loss owing to the size of the aperture used for the extraction of the spectra (see Appendix \ref{appendix_observations} for details). 

\begin{table}[h!]
    \centering
    \begin{tabular}{llll}
    \toprule
         Parameter  & Component & Prior type   & Prior range\\
         \midrule
         Age [Gyr]            & SSP & uniform  & [0,15]\\
         $Z/{Z_\odot}$ & SSP & uniform  & [0.01, 3]\\
         $\tau$ [Gyr]         & general & uniform  & [0,0.8]\\
         $\mass$     & general & uniform  & [6, 15]\\
         $\sigma_\star$ [km/s] & general &  $\mathcal{G}$(val, err) & [0, 400]\\
         $A_{\rm{V}}$ [mag] & dust & uniform  & [0, 4]  \\
         $n$                    & noise & uniform  & [1, 10] \\
         $C_i$ \quad \makecell[l]{$i$=0 \\ $i$=1,2} & calibration &
         \makecell[l]{$\mathcal{G}$(1,0.1) \\ $\mathcal{G}$(0,0.1)} &
         \makecell[l]{\([0.75, 1.25]\) \\ \([-0.25, 0.25]\)} \\
    \bottomrule
    \end{tabular}
    \caption{List of parameters, priors, and prior ranges. The $\mathcal{G}(x,y)$ stands for a Gaussian prior centered in $x$ with dispersion $y$. In the case of $\sigma_{\star}$, val and err refer to each galaxy's velocity dispersion and associated error, measured as described in Sect. \ref{sec:2DATA}.}
    \label{tab:priors}
\end{table}

In summary, the model was built on a set of ten parameters: age, mass ($\mass$), metallicity ($Z/Z_\odot$), stellar velocity dispersion ($\sigma_\star$), SFH width ($\tau$), dust reddening ($A_{\rm{V}}$), noise parameter ($n$), and the three coefficients of the calibration polynomial ($C_i$, $i$=0,1,2). The priors adopted for each parameter are presented in Table \ref{tab:priors}. 

Differently from what is common practice in FSF codes, no cosmological prior was adopted on the age parameter, allowed to span the range 0 -- 15 Gyr independently of redshift. This modification of the code, vastly tested and validated in previous works \citep{Jiao2023, Tomasetti2023, tomasetti_globular_2025}, is fundamental to ensure that the final result is independent of any cosmological model. 
\begin{figure}[h!]
    \centering
    \includegraphics[width=0.98\linewidth]{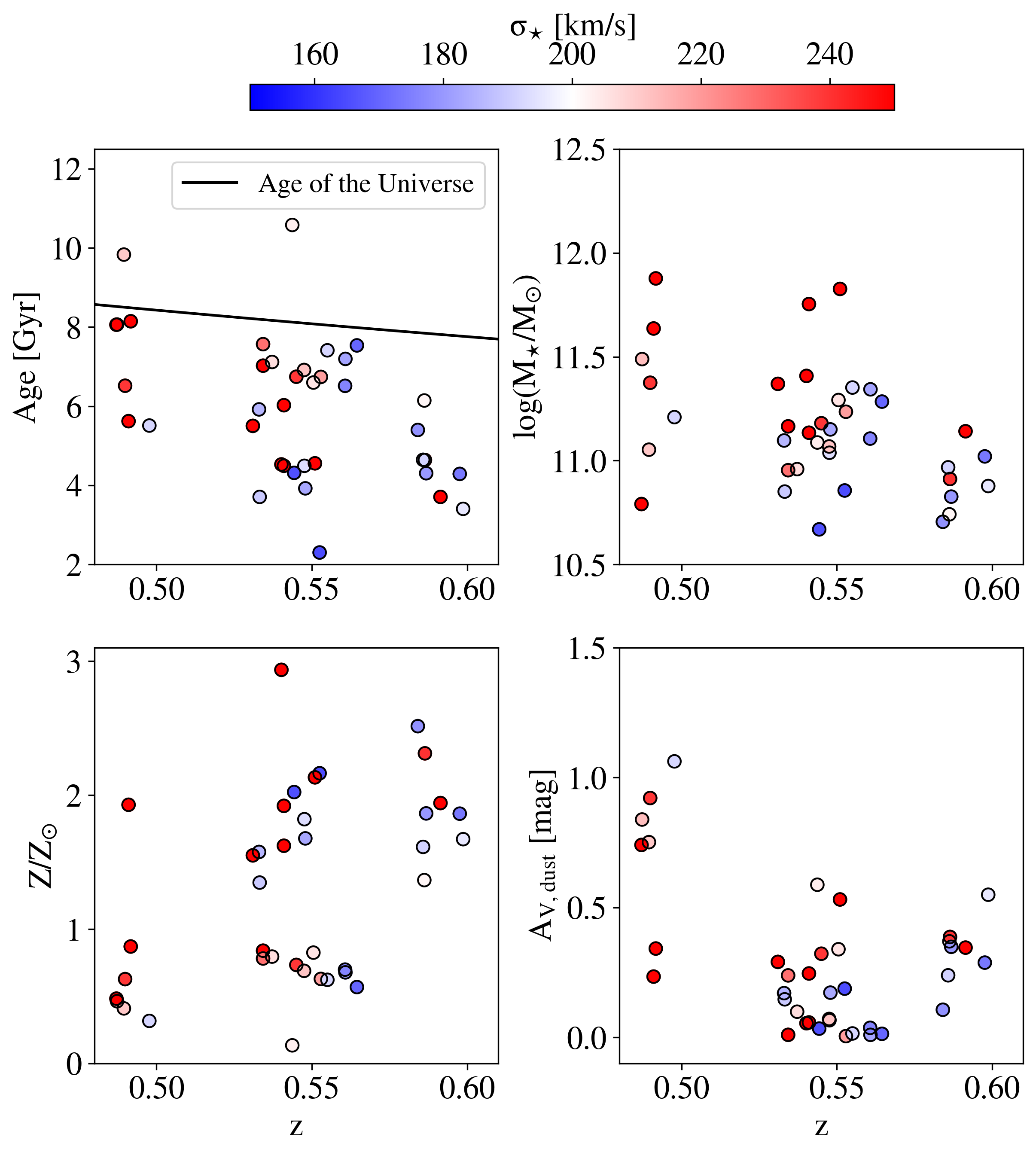}
    \caption{Trends in redshift for the main physical parameters, colour-coded by the measured velocity dispersion. In the top left panel, the solid line shows the age of the Universe computed in a flat $\Lambda$CDM with $\Omega_m=0.3$.}
    \label{fig:var_z}
\end{figure}

Before performing the fit, we masked the spectral regions potentially affected by telluric absorption or sky emission lines. In the observed frame, we excluded a 20 {\AA} wide window around the sky lines at [5577.338, 5889.959, 6300.0, 6363.0, 6863.955, 7964.65, 7993.332] {\AA}. We also masked the telluric absorption bands in the intervals [3000–3800] {\AA}, [6860–7000] {\AA}, and [7170–7350] {\AA}. Additionally, although these galaxies do not exhibit detectable emission lines, the regions of potential emission were also masked. In the rest frame, we excluded windows with a width corresponding to 2000 $\rm{km\: s^{-1}}$, centred at [3727, 4341, 4861.3, 4958.9, 5006.9, 6300, 6363, 6548, 6583, 6716, 6731] {\AA}.

\subsection{Results}
After the FSF, a visual inspection of the results was performed for each galaxy, to ensure a good convergence of the parameters and an accurate reproduction of the spectra and photometry. Only one object was flagged as a bad fit, owing to the presence of double peaks in the posterior probability distribution, and excluded from the sample, thus counting 37 CCs. The quality of the fit is also demonstrated by the average reduced chi-square values that, considering both spectrum and photometry, equals $\tilde{\chi}^2 = 1.6$. Accounting for the noise factor introduced in the fit, this further reduces to $\tilde{\chi}^2 = 0.54$. In Fig. \ref{fig:fit_example}, an example of a fit for each cluster is reported. 
\begin{figure*}
    \centering
    \includegraphics[width=0.95\linewidth]{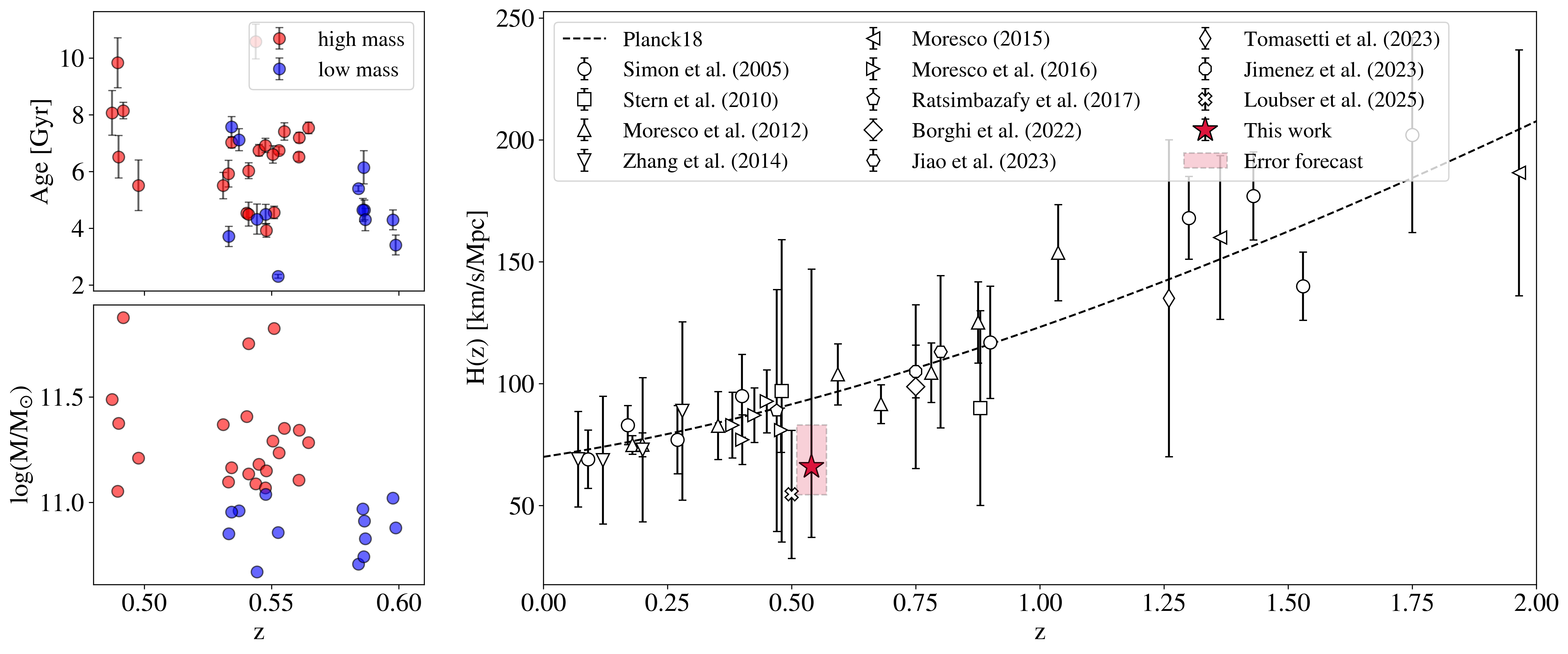}
    \caption{Collection of all $H(z)$ measurements obtained to date \citep{simon_constraints_2005,stern_cosmic_2010,Moresco2012,zhang_four_2014,Moresco2015,Moresco2016,ratsimbazafy_age-dating_2017,Borghi2022b,Jiao2023,Tomasetti2023,Jiao2023,loubser_independent_2025,loubser_measuring_2025}, including the result of this work. For comparison, the dashed line represents the flat $\Lambda$CDM trend from Planck18. The red shaded box shows the forecast on the precision achievable with a sample of 100 CCs (see Sect. \ref{sec:future_persp} for details).}
    \label{fig:H(z)}
\end{figure*}

Fig. \ref{fig:var_z} illustrates the redshift dependence of the main physical parameters: age, mass, metallicity, and dust extinction, colour-coded by stellar velocity dispersion ($\sigma_{\star}$). The CC sample is characterized by high stellar masses, with a mean of $\log(M_{\star}/M_\odot) = 11.2 \pm 0.3$ and, as expected, stellar mass correlates positively with velocity dispersion, such that the most massive galaxies also exhibit the highest $\sigma_{\star}$. The mass distribution is not uniform in redshift: higher-mass galaxies dominate the low-redshift end ($z\leq$ 0.55), whereas lower-mass systems become more prevalent at higher redshifts. This must be taken into account in the cosmological analysis of the age–redshift relation to ensure homogeneity across the subsamples used to trace population ageing.\\ 
A qualitative inspection of the age–redshift trend already highlights two key results: evidence of mass downsizing \citep{Thomas2010, Citro2017}, with high-$\sigma_\star$ galaxies appearing systematically older than their low-$\sigma_\star$ counterparts, and a clear overall ageing trend of the CC sample, with stellar ages decreasing steadily towards higher redshift in both the high- and low-$\sigma_\star$ populations. Even though no cosmological prior was adopted in the recovery of the galaxies' ages, 95\% of the sample is compatible with the age of the Universe in a flat $\Lambda$CDM ($\Omega_m=0.3$) within errors ($\sim 0.4$ Gyr on average). All objects are also characterized by short bursts of star formation, demonstrated by low values of the $\tau$ parameter, on average $\mathrm{\tau} = 0.6 \pm 0.2$ Gyr. We find typically super-solar metallicities, with a mean value of $Z/Z_{\odot} = 1.3 \pm 0.7$, but the sample spans the whole range 0--3 $Z_{\odot}$. Dust reddening is predominantly low, with an average extinction in V-band $A_{\rm{V}} = 0.3 \pm 0.3$ mag.

\section{Application to cosmology}
\label{sec:4COSMO}

Once the age--redshift relation for the sample of chronometers is derived, the CC method allows one to directly measure the Hubble parameter $H(z)$ by evaluating the slope of this trend. In particular, the expansion rate can be expressed as
\begin{equation}
H(z) = -\frac{1}{1+z}\,\frac{\mathrm{d}z}{\mathrm{d}t},
\label{eq:Hofz}
\end{equation}
where d$z$ is the redshift difference between two adjacent bins and d$t$ is the corresponding differential age. In practice, this requires a careful estimate of the CCs' differential ages, and a quantification of the associated statistical and systematic uncertainties.  

In this work, analysing three galaxy clusters, we fixed the redshift of each galaxy to that of its parent cluster, namely $z = 0.489$, $0.542$, and $0.588$ (bins 0, 1, and 2, respectively).
To evaluate the ageing of galaxies with comparable stellar masses, we divided the sample into two mass bins. As a threshold, we adopted $\mass = 11.05$, which minimises the difference between the mean stellar masses of analogue subsamples. As already noted in the previous section, the mass distribution is not homogeneous: considering bins 0, 1, and 2, the high-mass (HM, $\mass > 11.05$) subsample contains 5, 17, and 1 CCs, while the low-mass (LM, $\mass \leq 11.05$) subsample includes 1, 6, and 7 galaxies, respectively. Counting only one galaxy, both the last HM bin (HM2) and the first LM bin (LM0) had to be discarded. For this reason, we could evaluate the ageing of only two pairs of subsamples, HM0--HM1 and LM1--LM2, represented in the left panels of Fig. \ref{fig:H(z)}, to be combined into a single $H(z)$ measurement.

To this end, we decided to adopt a bootstrap approach, described in detail in Appendix \ref{appendix_boots}. We generated $N$=1000 perturbed realisations of the galaxy ages by sampling from Gaussian distributions centred on the best-fit ages and with a sigma equal to the age errors. For each realization, we performed $K$=1000 bootstrap resamplings, computed the median ages in each mass–redshift bin, and derived age differences d$t$ between adjacent bins. Converting those into $H(z)$ via Eq.~\ref{eq:Hofz} and averaging HM and LM results, we obtained $N$ bootstrap realizations of $H(z)$. The final value of $H(z)$ and its statistical uncertainty are represented by the median, $\mathrm{16^{th}}$ and $\mathrm{84^{th}}$ percentiles of this distribution:
\begin{align*}
    H(z=0.542) = 66_{-29}^{+81}\: \text{(stat) km s $^{-1}$ Mpc$^{-1}$}.
\end{align*}

\subsection{Systematic effects}
With only 35 CCs, the statistical uncertainties are inevitably large and strongly dominate over systematics. Nevertheless, we want to include an estimate of systematic effects, following the results in \citet{Tomasetti2023}, where the CC method was applied with the same FSF approach and a comparable sample size (39 CCs). In particular, the impact of the SFH was considered, and found to affect $H(z)$ at the $\sim$20\% level. We stress, however, that this value likely overestimates this systematic contribution, as the effect of SFH fluctuations would be significantly reduced in larger samples. Considering this 20\% as a conservative estimate of the systematics at play, and adding it in quadrature to the statistical component, we find:
\begin{align*}
    H(z=0.542) &= 66_{-29}^{+81}\: \text{(stat)} \pm 13\: \text{(syst)}\:\text{km s$^{-1}$ Mpc$^{-1}$}\\
    &=66_{-32}^{+82} \text{ (stat+syst)} \:\text{km s$^{-1}$ Mpc$^{-1}$}
\end{align*}

\subsection{Future perspectives}\label{sec:future_persp}
As current and future facilities (e.g., \textit{Euclid}, \citealt{Laureijs2011}; WST, \citealt{bacon_wst_2024}) will deliver spectroscopic and photometric data for vast galaxy samples, it is timely to explore the potential gains from larger statistics. To this end, we estimated the precision achievable on $H(z)$ with samples of different sizes, or different distributions, drawing synthetic age samples based on a flat $\Lambda$CDM cosmology, and applied the same bootstrap procedure described at the beginning of this section to retrieve the $H(z)$ measurement. In order to compare the results to this work, we considered the same division into two HM and two LM bins, four in total. 

We performed various types of simulations, varying the distribution in redshift, the redshift coverage, and the statistics of the CC sample, to forecast the kind of statistical uncertainty achievable in future works. These simulations followed the bootstrap procedure described in Appendix \ref{appendix_boots}, with one modification: at step A of the algorithm, we generated random samples drawn from a Gaussian distribution centered on the age predicted by a flat $\Lambda$CDM cosmology for an object formed at redshift $z_\mathrm{f}=3$ (for the HM sample) and $z_\mathrm{f}=1.5$ (for the LM sample), with a dispersion of 0.5 Gyr.

\begin{figure}
    \centering
    \includegraphics[width=\linewidth]{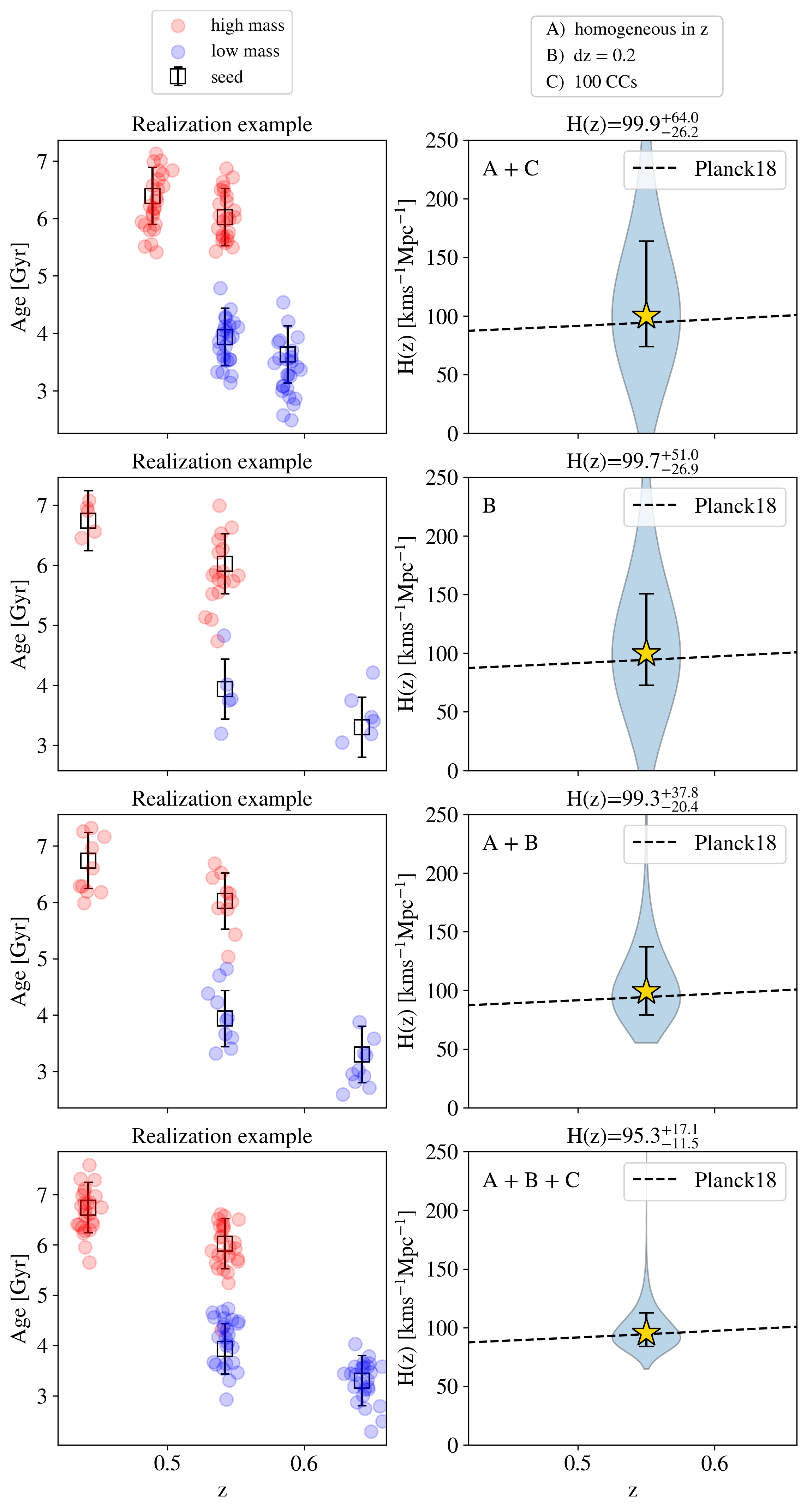}
    \caption{Results of the simulations. The left column shows, for each setting, an example of age--redshift both for the HM (red) and LM (blue) samples, randomly extracted from a Gaussian distribution centred on the \textit{seed}, in black. The right column shows the resulting $H(z)$ distribution and the derived measurement (yellow star, also reported in the title), in comparison with the assumed cosmology \citep{PlanckCollaboration2020}.}
    \label{fig:H_z_simulations}
\end{figure}

In this work, we analysed a sample of 35 CCs, unevenly distributed in four bins (with 5, 17, 6, and 7 galaxies respectively), covering a redshift interval d$z=0.1$. In these simulations, we tested three variations to this baseline: A) homogenizing the sample in redshift, so that each bin has the same number (10) of CCs, B) enlarging the redshift interval from d$z=0.1$ to d$z=0.2$, and C) increasing the sample size to 100 CCs.
In Fig. \ref{fig:H_z_simulations}, we show the results of some of the tests performed, including different combinations of the three modifications listed above. 

Interestingly, extending the redshift interval alone yielded the largest gain, with an error on $H(z)$, compared to the one found in this work, reduced by $\sim$30\%. Increasing the sample size also had a significant, though smaller, effect, reducing the error by $\sim$20\%. Homogenizing the distribution across bins had little impact on its own, but in combination with a wider redshift interval, it reduced the error by nearly 50\%. As expected, combining all three improvements provided the strongest result, lowering the uncertainty on $H(z)$ by approximately a factor of 4.

\section{Conclusions}\label{sec:5CONCLUSIONS}

In this work, we applied the CC method to measure the expansion history of the Universe, analysing VLT/MUSE data for the most massive, passively evolving galaxies in three galaxy clusters: SDSS 2222 ($z= 0.49$), MACS 1149 ($z = 0.54$), and SDSS 1029 ($z= 0.59$). Since MACS 1149 also hosts the multiply imaged SN Refsdal, previously used to infer $H_0$ via TDC, our measurement of $H(z)$ from the ageing of its member galaxies enables a direct combination of complementary probes, thereby strengthening constraints on cosmological parameters.

We first selected the CC sample by combining a cut in stellar velocity dispersion, on their \textit{S/N}, and on the H/K ratio, to isolate the most massive and passive cluster members, ending up with a sample of 38 CCs. We then performed FSF with the \texttt{Bagpipes} code, specifically modified to remove the cosmological prior on ages, so that they could vary up to 15 Gyr. Despite this, 95\% of the sample results compatible with the age of the Universe in a flat $\Lambda$CDM model, and shows an ageing trend with redshift. Concerning the other physical parameters, the CC sample shows the characteristics of a passive population, with high stellar mass ($\log(M_{\star}/M_\odot) = 11.2 \pm 0.3$), low dust extinction ($A_{\rm{V}} = 0.3 \pm 0.3$ mag), and short timescale of formation ($\tau = 0.6 \pm 0.2$ Gyr). The metallicity is on average super-solar, with a mean of $Z/Z_{\odot} = 1.3 \pm 0.7$.

From the age--redshift relation, separating the high-mass and the low-mass galaxies (with $\mass=11.05$ as a threshold to ensure maximum homogeneity among subsamples), we measured $H(z)$ applying the CC method through a bootstrap approach, obtaining: $H$($z=0.542$) = $66_{-29}^{+81}$ (stat) $\pm 13$ (syst). The error budget is currently dominated by the statistical component, but simulations showed that with $\sim$100 CCs and an extended redshift interval of 
d$z=0.2$, the statistical uncertainty could be reduced by up to 75\%. Such an increase in precision would provide sufficiently tight constraints on $H(z)$ that, when combined with TDC measurements, could significantly enhance their joint constraining power on cosmological parameters by breaking existing degeneracies (e.g., the $H_0$–$\Omega_m$ degeneracy; see \citealt{Bergamini2024}). 

Future and upcoming surveys (e.g., \textit{Euclid}, WST, MOONS, LSST) will provide large samples of galaxies and galaxy clusters, offering unprecedented opportunities to apply the TDC method to strong lensing galaxy clusters, and the CC method to those cluster members with the required spectroscopic quality (or the necessary follow-up observations). \textit{Euclid} \citep{Laureijs2011}, in particular, is expected to increase the census of strong gravitational lenses by orders of magnitude \citep{euclid_collaboration_euclid_2025}, opening the door to a systematic exploitation of this synergy.

\begin{acknowledgements}
 The authors warmly thank Ana Acebron and Eros Vanzella for their useful and constructive feedback to this work. ET acknowledges the support from COST Action CA21136 – “Addressing observational tensions in cosmology with systematics and fundamental physics (CosmoVerse)”, supported by COST (European Cooperation in Science and Technology). MMo acknowledges the financial contribution from the grant PRIN-MUR 2022 2022NY2ZRS 001 “Optimizing the extraction of cosmological information from Large Scale Structure analysis in view of the next large spectroscopic surveys” supported by Next Generation EU. MMo and AC acknowledge support from the grant ASI n. 2024-10-HH.0 “Attività scientifiche per la missione \textit{Euclid} – fase E”. The authors acknowledge support from the Italian Ministry of University and Research through grant PRIN-MIUR 2020SKSTHZ. MMe and PB were supported by INAF Grants “The Big-Data era of cluster lensing" and "Probing Dark Matter and Galaxy Formation in Galaxy Clusters through Strong Gravitational Lensing", and ASI Grant n. 2024-10-HH.0 "Attività scientifiche per la missione Euclid – fase E". The research activities described in this paper have been co-funded by the European Union – NextGeneration EU within PRIN 2022 project no. 20229YBSAN – Globular clusters in cosmological simulations and in lensed fields: from their birth to the present epoch. SS has received funding from the European Union’s Horizon 2022 research and innovation programme under the Marie Skłodowska-Curie grant agreement No 101105167 — FASTIDIoUS.
    
\end{acknowledgements}

\bibliographystyle{aa}
\bibliography{references}
\newpage

\begin{appendix}

\section{Observations}\label{appendix_observations}

\begin{figure*}[t!]
    \centering
    \includegraphics[width=0.8\linewidth]{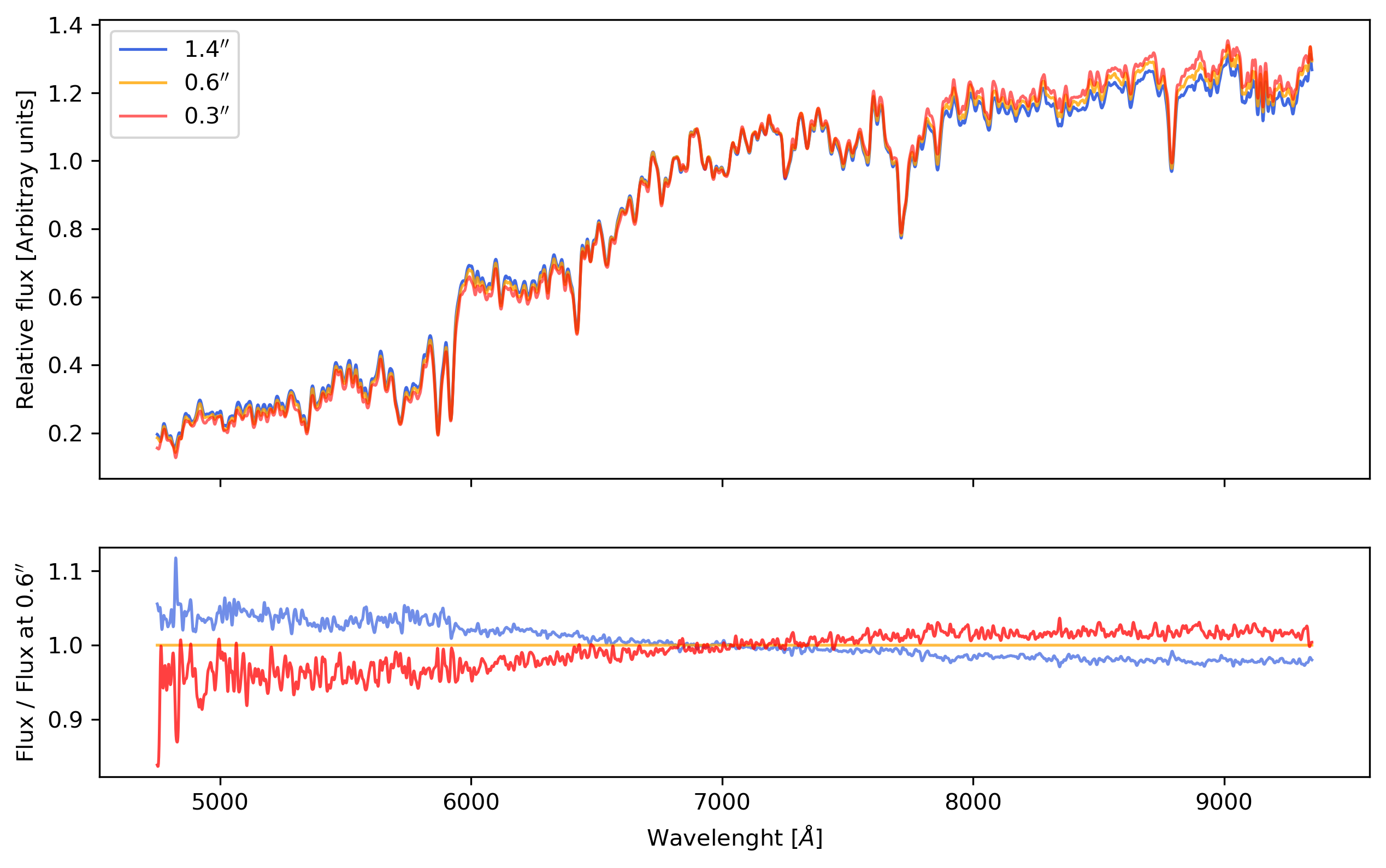}
    \caption{Example showing the effect of the aperture sizes on the extraction of a 1D spectrum from a MUSE data cube. The top panel shows the spectrum of the same member of the cluster SDSS 2222, extracted using circular apertures of radius $1.4''$ (blue), $0.6''$ (orange), and $0.3''$ (red). The bottom panel shows the ratio between each spectrum and the one extracted with the $0.3''$ aperture. A bigger extraction aperture produces a relatively bluer spectrum (namely, the spectrum shows a larger flux at lower wavelengths).}
    \label{fig:spec_ap_test}
\end{figure*}

MUSE coverage of the central regions of MACS 1149 was granted by the DDT programme 294.A-5032 (P.I. C. Grillo), for a total integration time of 6h and seeing of less than $1.1''$ in almost all exposures \citep{Grillo2016}, complemented by $5.5 \, \mathrm{h}$ of observations of the north-west region of the cluster within the programme 105.20P5 (P.I. A. Mercurio), presented in \citet{Schuldt2024}. SDSS 1029 was observed with MUSE under the programme 0102.A-0642(A) (P.I. C. Grillo), presented in \citet{acebron_vltmuse_2022}, with a total exposure time of 4.8 h and a FWHM of $0.71''$. SDSS 2222 was targeted with the integral field spectrograph MUSE at the VLT, under programme 0103.A-0554(A) (P.I. C. Grillo), presented in \citet{acebron_new_2022} for a cumulative exposure time of 4.4 h. We used the MUSE data-reduction pipeline version \texttt{v2.8.5} \citep{Weilbacher20} for all observations to build the final data cube.

We used archival \textit{Hubble} Space Telescope (HST) multicolour imaging from the Advanced Camera for Surveys (ACS) and the Wide Field Camera 3 (WFC3). SDSS 2222 (GO-13337; P.I. Sharon) was imaged over two orbits in each of the ACS filters F475W, F606W, and F814W, while the WFC3 imaging in F110W and F160W was allocated one single orbit in total. SDSS 1029 (GO-12195; P.I. Oguri) was imaged over two orbits in F475W, three orbits in F814W, and two orbits in the WFC3 filter F160W. A detailed description of the observations and data reduction process is available in \citet{Sharon2017, Oguri2013}. For MACS 1149, the photometric data belong to the Frontier Fields program
\citep[P.I. J. Lotz,][]{lotz_frontier_2017}, while the data reduction is widely described in \citet{tortorelli_kormendy_2023}. For each cluster, we extracted a photometric catalogue using \texttt{Morphofit}\footnote{\texttt{Morphofit}: a Python package, based on Galfit \citep{Peng2002,Peng2010} and Sextractor \citep{Bertin1996}, for the morphological analysis of galaxies. See \url{https://github.com/torluca/morphofit}}~\citep{Tortorelli2023a}. We also investigated how the choice of aperture size affects the extraction of 1D spectra from MUSE data cubes. For a subset of galaxies, we extracted their spectra from the data cube using circular apertures of $0.5''$, $0.8''$, $1.0''$, and $1.6''$ of radius. As shown in Fig. \ref{fig:spec_ap_test}, we found that the shape of the 1D spectrum varies with the aperture size. Specifically, a smaller extraction aperture tends to produce a redder spectrum compared to one extracted using a larger aperture (namely, the spectrum shows an increase in flux at longer wavelengths). This is not unexpected, since often early type galaxies show colour gradients due to, for example, a lower metallicity of the stars inhabiting the galaxy outskirts compared to the ones that are in the central region \citep[e.g.][]{Peletier1990,LaBarbera2005}. This effect can cause a non-negligible mismatch between the galaxy spectrum and the photometry when different apertures are used, negatively impacting the quality of the FSF. For this reason, we also allowed \texttt{Bagpipes} to perform a calibration of the spectrum by multiplying it with a second-order Chebyshev polynomial and, with the coefficients optimized during the fitting process, to achieve a better match of the spectrum with the photometry.

\section{Bootstrap algorithm}\label{appendix_boots}
\begin{figure}[h!]
    \centering
    \includegraphics[width=0.95\linewidth]{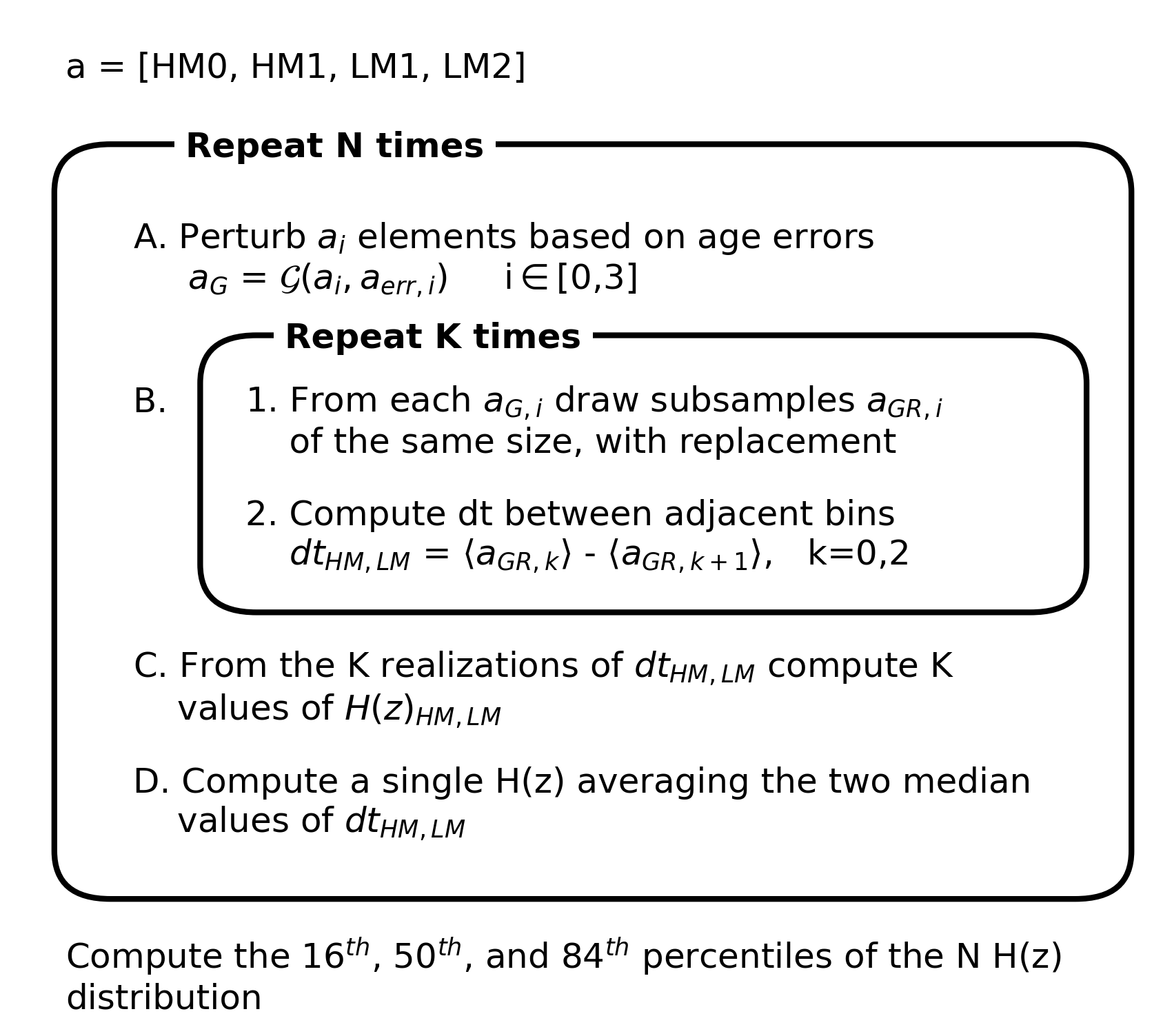}
    \caption{Visual diagram of the bootstrap algorithm. The CC sample was divided into three redshift bins (0, 1, 2) and into two mass bins, high-mass (HM) and low-mass (LM), but only bins HM0, HM1, LM1, and LM2 have enough statistics to be used.}
    \label{fig:boots}
\end{figure}

In Fig. \ref{fig:boots}, the workflow of the bootstrap algorithm used to obtain the final $H(z)$ measurement is illustrated. Each cycle is repeated a thousand times ($N=K=1000$). 

We generated $N=1000$ perturbed realizations of the galaxy ages by sampling each object from a Gaussian distribution centred on its best-fit age and with a standard deviation equal to its uncertainty. From each of these perturbed arrays, we extracted with repetition $K=1000$ samples with the same numerosity, computed the $K$ median ages in each mass--redshift bin, and estimated the age differences d$t$ between adjacent bins of the same mass. This process generated $N$ bootstrap distributions of d$t$, each with $K$ realisations, for both the high- and the low-redshift intervals. For each case, we extracted the median of d$t$, derived two independent measurements of $H(z)$ applying Eq.~\ref{eq:Hofz}, and combined them into a single measurement by averaging the values. The final value of $H(z)$ and its statistical uncertainty are represented by the median, $\mathrm{16^{th}}$ and $\mathrm{84^{th}}$ percentiles of these $N$ realizations.

\end{appendix}

\end{document}